\documentclass[pra,twocolumn,tightenlines,showpacs,nofootinbib]{revtex4}
\usepackage{bm,dcolumn,amsmath,graphicx}
\usepackage{epsfig}

\newcommand{\tref}[1]{Table~\ref{#1}}

\newcommand{\TaV}{Ta$^{4+}$}
\newcommand{\WVI}{W$^{5+}$}
\newcommand{\ReVII}{Re$^{6+}$}
\newcommand{\AcIII}{Ac$^{2+}$}
\newcommand{\UVI}{U$^{5+}$}
\newcommand{\ZrIII}{Zr$^{2+}$}
\newcommand{\HfIII}{Hf$^{2+}$}
\newcommand{\AcII}{Ac$^{+}$}
\newcommand{\HgIII}{Hg$^{2+}$}
\newcommand{\HgIV}{Hg$^{3+}$}

\begin{document}
\title{Transitions in Zr, Hf, Ta, W, Re, Hg, Ac and U ions with high
  sensitivity to variation of the fine-structure constant}

\author{J. C. Berengut}
\author{V. A. Dzuba}
\author{V. V. Flambaum}
\affiliation{School of Physics, University of New South Wales,
                  Sydney, NSW 2052, Australia}
 
\date{22 September 2011}

\begin{abstract}

We study transitions between ground and low-energy excited states of heavy
ions corresponding to $s-d$ single-electron transitions or
$s^2-d^2$ double-electron transitions. The large nuclear charge $Z$ and
significant change in angular momentum of electron orbitals make these
transitions highly sensitive to a potential variation in the fine-structure constant, $\alpha$.
The transitions may be considered as candidates for laboratory searches for
space-time variation of $\alpha$. 
%As a byproduct of this work we have calculated the energies
%of the lowest states of ions for which experimental data is not
%available or controversial (e.g, \HgIV).  

\end{abstract}

\maketitle

%================================================================
\section{Introduction}
%================================================================
Theories unifying gravity with other interactions suggest that
the fundamental constants of nature may vary over space and time
(see, e.g., the review~\cite{Uzan}).
Indications that the fine-structure constant,
$\alpha = e^2/\hbar c$, might change over cosmological scales
has been found in quasar absorption spectra~\cite{Webb99,Webb01,Murphy01a,Murphy01b,Murphy01c,Murphy01d}.
The most recent analysis of around 300 quasar absorption spectra taken
from Keck and VLT telescopes is consistent with a smooth spatial gradient in
the values of $\alpha$ along a particular direction in space~\cite{dipole}.
This ``Australian dipole'' reconciles all existing astrophysical and 
laboratory (null) measurements~(\cite{berengut11jpcs}, see also~\cite{berengut11prd}).

Since the solar system (and the Earth within it) moves with respect to the
frame of the gradient, the values of $\alpha$ on Earth should change as we
move from regions of space with smaller $\alpha$ to regions with larger values.
Thus the spatial variation of $\alpha$ may be studied using
laboratory experiments that measure the change of $\alpha$ in
time~\cite{BF10}. A number of such experiments have already been performed and
reported (see, e.g. the review~\cite{FD09}). The best current limit, 
$\dot\alpha/\alpha = (-1.6 \pm 2.3) \times 10^{-17}\ \textrm{year}^{-1}$, comes
from comparison of Hg$^+$ and Al$^+$ optical clocks over the course of
a year~\cite{Rosenband}. This limit is better than that obtained
from quasar absorption spectra if one assumes a linear time varation of $\alpha$
over $\sim10^{10}$~year timescales.
However, it needs to be further improved by two or three orders-of-magnitude
to test the Australian dipole hypothesis~\cite{BF10}.
One way of achieving this is to find an atomic system where the spectra
is significantly more sensitive to the change
of $\alpha$ than in the Hg$^+$/Al$^+$ system. 

In our first paper on this subject~\cite{DFW99} we suggested using
$s-d$ optical transitions in heavy atoms and ions. The sensitivity
of these transitions to variation of $\alpha$ is large. This is
exactly what was used in the Hg$^+$/Al$^+$ experiment~\cite{Rosenband}. 

A number of other atomic transitions have been found where sensitivity
of the frequencies of the transitions to the variation of the fine-structure
 constant is even higher. These include close, long-lived
states of different configurations~\cite{DFW99,ADF06,Dzuba08}, fine-structure
anomalies~\cite{DF05}, and optical transitions in highly charged
ions~\cite{BDF10,berengut11prl}. The near-degenerate excited states of
dysprosium have already been used to place strong limits on terrestrial
$\alpha$-variation~\cite{Budker04,Budker07,Budker07a}.

In present paper we further study the original idea of using the $s-d$
and $s-f$ transitions in heavy ions. We consider a number of ions
and find many transitions which are as good as that used in Hg$^+$,
or even better. Results for ions with one valence electron above closed shells
are presented in Sec.~\ref{sec:one_valence}; these are Tm-like \TaV, \WVI, and \ReVII,
and Fr-like \AcIII and \UVI. In Sec.~\ref{sec:two_valence} we study ions
with two electrons above closed shells: Sr-like \ZrIII, Yb-like \HfIII, and Ra-like \AcII.
Finally we consider the \HgIII\ and \HgIV\ mercury ions in Sec.~\ref{sec:closed_shells},
which have closed shells and single-hole electronic structures, respectively.
In all cases we have chosen ions with relatively low-energy transitions, so that
they may be within the range of lasers -- an important criteria for
potential clocks.

%================================================================
\section{Calculations}
%================================================================

The dependence of atomic frequencies on the fine-structure constant
appears due to relativistic corrections. In the
vicinity of its present laboratory value $\alpha_0 \approx 1/137$ it is
presented in the form 
\begin{equation}
  \omega(x) = \omega_0 + qx,
\label{omega}
\end{equation}
where $\omega_0$ is the laboratory value of the frequency and
$x = (\alpha/\alpha_0)^2-1$. $q$ is the sensitivity coefficient
which must be found from atomic calculations:
\begin{equation}
 q = \left .\frac{d\omega}{dx}\right|_{x=0}
  \approx  \frac{\omega(\delta x) - \omega(-\delta x)}{2\cdot\delta x}
\label{qq}
\end{equation}
Here $\delta x$ must be small enough to exclude terms non-linear in $\delta (\alpha^2)$,
yet should be large enough to ensure numerical stability.
In the present calculations we use $\delta x = 0.01$.

For all atoms we start the calculations from the relativistic
Hartree-Fock method (RHF). Techniques to include correlations depend
on atomic structure and we will discuss them in the relevant sections.

\subsection{Ta V, W VI, Re VII, Ac III, and U VI} 
\label{sec:one_valence}

\begin{table}
\caption{Energies and sensitivity coefficients ($q$) for isoelectronic
  sequence \TaV, \WVI\ and \ReVII\ (cm$^{-1}$).}
\label{T:TaRe}
\begin{ruledtabular}
\begin{tabular}{lccrrr}
\multicolumn{1}{c}{Ion} & 
\multicolumn{1}{c}{$Z$} & 
\multicolumn{1}{c}{Level} & 
\multicolumn{2}{c}{Energy} &
\multicolumn{1}{c}{$q$} \\
 & & & \multicolumn{1}{c}{Experiment}  & \multicolumn{1}{c}{Theory} \\
\hline
\TaV    & 73 & $5d_{3/2}$ &      0                 &      0 &       0  \\
        &    & $5d_{5/2}$ &   6608\footnotemark[1] &   5833 &    5161  \\
        &    & $6s_{1/2}$ &  47052\footnotemark[1] &  44812 &  -30931  \\
\WVI    & 74 & $5d_{3/2}$ &      0                 &      0 &       0  \\
        &    & $5d_{5/2}$ &   8707\footnotemark[2] &   7981 &    7292  \\
        &    & $6s_{1/2}$ &  79433\footnotemark[2] &  77293 &  -38423  \\
\ReVII  & 75 & $5d_{3/2}$ &      0                 &      0 &       0  \\
        &    & $5d_{5/2}$ &  10996\footnotemark[3] &  10410 &    9626  \\
        &    & $6s_{1/2}$ & 115066\footnotemark[3] & 112575 &  -46470  \\
\end{tabular}
\footnotetext[1]{Reference~\cite{Meijer73}}
\footnotetext[2]{Reference~\cite{Meijer74}}
\footnotetext[3]{Reference~\cite{Meijer76}}
\end{ruledtabular}
\end{table}

\begin{table}
\caption{Energy levels and sensitivity coefficients ($q$) for \AcIII\
  (cm$^{-1}$).}
\label{T:AcIII}
\begin{ruledtabular}
\begin{tabular}{lrrr}
\multicolumn{1}{c}{Level} &
\multicolumn{2}{c}{Energy} &
\multicolumn{1}{c}{$q$} \\
 & \multicolumn{1}{c}{Experiment\tablenotemark[1]}  & \multicolumn{1}{c}{Theory} \\
\hline
$7s_{1/2}$ &      0 &      0 &        0  \\
$6d_{3/2}$ &    801 &   1582 &    27297  \\
$6d_{5/2}$ &   4204 &   5449 &    30230  \\
$5f_{5/2}$ &  23455 &  21698 &    56170  \\
$5f_{7/2}$ &  26080 &  24845 &    57324  \\
$7p_{1/2}$ &  29466 &  30542 &     6861  \\
$7p_{3/2}$ &  38063 &  39550 &    19118  \\
\end{tabular}
\footnotetext[1]{Reference~\cite{Moore}}
\end{ruledtabular}
\end{table}

\begin{table}
\caption{Calculated energy levels and sensitivity coefficients ($q$) for \UVI\
  (cm$^{-1}$).}
\label{T:UVII}
\begin{ruledtabular}
\begin{tabular}{lrr}
\multicolumn{1}{c}{Level} & 
\multicolumn{1}{c}{Energy} & 
\multicolumn{1}{c}{$q$} \\
\hline
$5f_{5/2}$ &      0 &       0  \\
$5f_{7/2}$ &   6960 &    4687  \\
$6d_{3/2}$ &  76173 &  -52900  \\
$6d_{5/2}$ &  84683 &  -46300  \\
$7s_{1/2}$ & 123368 & -110355  \\
$7p_{1/2}$ & 173761 &         \\
$7p_{3/2}$ & 195351 &         \\
\end{tabular}
\end{ruledtabular}
\end{table}

These ions have one external electron above closed shells. We use the
correlation potential method~\cite{DFSS87} in the $V^{N-1}$
approximation to perform the calculations. Initial Hartree-Fock
procedure is done for a closed-shell ion, with the external electron
removed. States of the external electron are calculated in the field
of frozen core. Correlations are included with the use of the
second-order correlation potential $\hat \Sigma$. 

We use the $B$-spline technique~\cite{B-spline} to generate a complete
set of single-electron states which are needed for the calculation
of $\hat\Sigma$. These states are the eigenstates of the
RHF Hamiltonian $\hat H_0$ with the $\hat V^{N-1}$ electron
potential. We use 50 $B$-splines of order 9 in a cavity of radius
40$a_B$. 

Energies for the valence states ($\epsilon_v$) are found by solving
the Brueckner orbital equations for external electron
\begin{equation}
  (\hat H_0 + \hat \Sigma - \epsilon_v) \psi_v = 0.
\label{eq:1el}
\end{equation}

The results for Yb$^+$-like tantalum, tungsten and rhenium and for Fr-like
actinium and uranium are presented in Tables~\ref{T:TaRe},
\ref{T:AcIII} and~\ref{T:UVII}.
All these ions have $s-d$ or $s-f$ transitions with large $q$
coefficients. Most of the states with large $q$-coefficients are
metastable states. Large $q$-coefficients can be either positive or
negative which can be used to further improve the sensitivity of the
frequency shift measurements to the time-variation of $\alpha$. One of
the most interesting system is \UVI, where the $q$-coefficient
for the frequency of the transition from the ground $5f_{5/2}$ state
to the metastable $7s$ state is very large ($q \sim -10^5$) and
negative. The ratio of this frequency to the fine structure interval
in the ground state (where $q$ is positive) is very sensitive to the
variation of $\alpha$. While such high energies are outside the range
of normal optical lasers, they can potentially be reached using high-UV
lasers, such as those that employ high-harmonic generation~\cite{cingoz11arxiv}.
\tref{T:UVII} also shows calculated
energies of the $7p_{1/2}$ and $7p_{3/2}$ states. It is useful to know
their positions to make sure that the $6d$ and $7s$ states are metastable. 
To the best of our knowledge the experimental
data on the energy levels of \UVI\ is absent. 

\subsection{Zr III, Hf III, and Ac II}
\label{sec:two_valence}

\begin{table}
\caption{Energy levels and sensitivity coefficients ($q$) for \ZrIII\
  (cm$^{-1}$).}
\label{T:ZrIII}
\begin{ruledtabular}
\begin{tabular}{lcrrr}
\multicolumn{2}{c}{Level} &
\multicolumn{2}{c}{Energy} &
\multicolumn{1}{c}{$q$} \\
 & & \multicolumn{1}{c}{Experiment\tablenotemark[1]}  & \multicolumn{1}{c}{Theory} \\
\hline
$4d^2$ & $^3$F$_2$ &      0 &      0 &        0  \\
$4d5s$ & $^3$D$_1$ &  18399 &  18554 &    -6018  \\
       & $^3$D$_2$ &  18803 &  18988 &    -5637  \\
       & $^3$D$_3$ &  19533 &  19764 &    -4860  \\
$4d5s$ & $^1$D$_2$ &  25066 &  26188 &    -4025  \\
$5s^2$ & $^1$S$_0$ &  48507 &  49297 &   -11000  \\
\end{tabular}
\footnotetext[1]{Reference~\cite{NIST}}
\end{ruledtabular}
\end{table}

\begin{table}
\caption{Energy levels and sensitivity coefficients ($q$) for \HfIII\
  (cm$^{-1}$).}
\label{T:HfIII}
\begin{ruledtabular}
\begin{tabular}{lcrr}
\multicolumn{1}{c}{Leading} & \multicolumn{1}{c}{Level} &
\multicolumn{1}{c}{Energy}  & \multicolumn{1}{c}{$q$} \\
\multicolumn{1}{c}{configurations} &&& \\
\hline
$5d^2  $ & $^3$F$_2$ &      0 &        0  \\
         & $^3$F$_3$ &   3558 &     5200  \\
         & $^3$F$_4$ &   6652 &     7700  \\
$5d6s  $ & $^3$D$_1$ &   2652 &   -19600  \\
         & $^3$D$_2$ &   3121 &   -14600  \\
         & $^3$D$_3$ &   7112 &   -15500  \\
55\% $5d6s  $ & $^1$D$_2$ &   6124 &    -6700  \\
68\% $5d^2$, 30\% $6s^2$ & $^3$P$_0$ &   8775 & -13500  \\
43\% $5d^2$, 55\% $6s^2$ & $^1$S$_0$ &  11403 & -19800  \\
$5d^2  $ & $^3$P$_1$ &  11345 &    5100  \\
         & $^3$P$_2$ &  12995 &    4400  \\

$5d^2  $ & $^1$G$_4$ &  16151 &    6200  \\

79\% $5d^2$, 13\% $6s^2$ & $^1$S$_0$ & 33745 & 400  \\
\end{tabular}
\end{ruledtabular}
\end{table}

\begin{table}
\caption{Energy levels and sensitivity coefficients ($q$) for \AcII\
  (cm$^{-1}$).}
\label{T:AcII}
\begin{ruledtabular}
\begin{tabular}{lcrrr}
\multicolumn{2}{c}{Level} &
\multicolumn{2}{c}{Energy} &
\multicolumn{1}{c}{$q$} \\
 & & \multicolumn{1}{c}{Experiment\tablenotemark[1]}  & \multicolumn{1}{c}{Theory} \\
\hline
$7s^2$ & $^1$S$_0$ &	     0 &    0 &    0   \\

$6d7s$ & $^3$D$_1$ &	  4740  &  5460 &  22640 \\
       & $^3$D$_2$ &	  5267  &  6083 &  22989 \\
       & $^3$D$_3$ &	  7427  &  8514 &  24269 \\
 	 	 	      
$6d7s$ & $^1$D$_2$ &	  9087  & 10385 &  30346 \\
 	 	 	      
$6d^2$ & $^3$F$_2$ &	 13236  & 14639 &  44520 \\
       & $^3$F$_3$ &	 14949  & 16552 &  47073 \\
       & $^3$F$_4$ &	 16757  & 18646 &  48178 \\
 	 	 	      
$6d^2$ & $^3$P$_0$ &	 17737  & 19204 & 44922  \\
       & $^3$P$_1$ &	 19015  & 20649 & 47038  \\
       & $^3$P$_2$ &	 22199  & 21132 & 45884  \\
\end{tabular}
\tablenotetext[1]{Reference~\cite{Moore}}
\end{ruledtabular}
\end{table}

These ions have two external electrons above closed shells. Energies
of the $s$ and $d$ valence states are close to each other which means
that states of the $s^2$ and $d^2$ configurations should be close as
well. Frequencies of the transitions between states of these
configurations are expected to be more sensitive to the variation of
$\alpha$ compared to the single-electron $s-d$ transitions.

For calculations we use the CI+MBPT method developed in our previous
works \cite{DFK96,DFK96a,DJ98,Dzu05,DF07}. Calculations are done in
the $V^{N-2}$ approximation with two valence electrons removed from
the initial Hartree-Fock procedure. Basis states for valence electrons
are calculated in the field of frozen core. Configuration interaction (CI)
technique is used to construct two-electron states of valence
electrons. Core-valence correlations are included by means of the
many-body perturbation theory (MBPT). The $B$-spline
technique~\cite{B-spline} is used to
calculate basis states for valence electrons and to calculate
core-valence correlation operator $\hat \Sigma$.

The results are presented in Tables~\ref{T:ZrIII}, \ref{T:HfIII} and~
\ref{T:AcII}. The results for \ZrIII\ and \AcII\ show that the
$q$-coefficients for the $d^2- s^2$ transition are indeed about two times
larger than for the $d - s$ transitions. It is natural to expect
larger values of $q$ for \HfIII\ than for \ZrIII\ due to larger $Z$. 
It turns out however, that
\HfIII\ has no states of pure $6s^2$ configuration: there is strong
mixing between the $6s^2$ and $5d^2$ configurations, and the weight of
the $6s^2$ configuration does not exceed 55\% (see
Table~\ref{T:HfIII}). This affects the values of $q$. They are not as
large as they would be for the pure $6s^2$ case. Note also that there
is no experimental data on the spectrum of \HfIII. 

The largest $q$-coefficients among these three ions are for the \AcII\
(Table~\ref{T:AcII}). This ion has excited metastable states of the
$6d^2$ configuration while the ground state is practically pure $7s^2$
configuration. 

\subsection{Hg III and Hg IV}
\label{sec:closed_shells}

\begin{table}
\caption{Energy levels and sensitivity coefficients ($q$) for \HgIII\
  (cm$^{-1}$).}
\label{T:HgIII}
\begin{ruledtabular}
\begin{tabular}{lcrrr}
\multicolumn{2}{c}{Level} &
\multicolumn{2}{c}{Energy} &
\multicolumn{1}{c}{$q$} \\
 & & \multicolumn{1}{c}{Experiment\tablenotemark[1]}  & \multicolumn{1}{c}{Theory} \\
\hline
$5d^{10}  $ & $^1$S$_0$ &      0 &      0 &        0  \\
$5d^9 6s  $ & $^3$D$_3$ &  42850 &  42191 &   -54800  \\
            & $^3$D$_2$ &  46029 &  45920 &   -52900  \\
            & $^3$D$_1$ &  58405 &  58500 &   -39600  \\
$5d^9 6s  $ & $^1$D$_2$ &  61085 &  61835 &   -39900  \\
$5d^8 6s^2$ & $^3$F$_4$ &  97893 &  97719 &  -114600  \\
\end{tabular}
\tablenotetext[1]{Reference~\cite{Moore}}
\end{ruledtabular}
\end{table}

\begin{table}
\caption{Energy levels and sensitivity coefficients ($q$)
  for \HgIV\ (cm$^{-1}$).}
\label{T:HgIV}
\begin{ruledtabular}
\begin{tabular}{lcrrr}
\multicolumn{2}{c}{Level} &
\multicolumn{2}{c}{Energy} &
\multicolumn{1}{c}{$q$} \\
 & & \multicolumn{1}{c}{Experiment\tablenotemark[1]}  & \multicolumn{1}{c}{Theory} \\
\hline
$5d^9$    & $^2$D$_{5/2}$ &     0 &      0 &    0   \\
          & $^2$D$_{3/2}$ & 15685 &  16140 &  14700 \\
                                           
$5d^8 6s$ & $^4$F$_{9/2}$ & 60138 &  59370 & -63100 \\
          & $^4$F$_{7/2}$ & 66109 &  66206 & -58600 \\
          & $^4$F$_{5/2}$ & 69942 &  71809 & -60900 \\
          & $^4$F$_{3/2}$ & 71763 &  73365 & -58500 \\
                                           
$5d^8 6s$ & $^2$F$_{7/2}$ & 78854 &  79805 & -47800 \\
          & $^2$F$_{5/2}$ & 77675 &  78337 & -48800 \\
                                           
$5d^8 6s$ & $^4$P$_{5/2}$ & 86031 &  88669 & -45700 \\
          & $^4$P$_{3/2}$ & 83916 &  86568 & -47600 \\
          & $^4$P$_{1/2}$ & 82391 &  86501 & -56100 \\
%          &              &        &        & \\
%......................................
$5d^7 6s^2$ & $^4$F$_{5/2}$ &    & 145120  &  -130300 \\
\end{tabular}
\tablenotetext[1]{References \cite{Joshi} and \cite{Valk}}
\end{ruledtabular}
\end{table}

Finally, we consider the \HgIII\ and \HgIV\ ions. Here additional enhancement
is expected due to excitations of the electrons from the (almost) filled
$5d$ subshell. Effective nuclear charge for electrons in almost filled
many-electron subshells are higher than for valence electrons outside
of the closed shells. Therefore, relativistic effects are larger and
the $q$-coefficients are larger too~\cite{berengut11prl}.

These ions have ten (\HgIII) and nine (\HgIV) external electrons. For
the calculations we use the method especially developed for the
many-electron cases in Refs.~\cite{Dzuba08a,Dzuba08b}. The results are
presented in Tables~\ref{T:HgIII} and~\ref{T:HgIV}. Here again we see
that the value of the $q$-coefficient for double $d-s$ transition is
about two times larger than for single $d-s$ transition. The value of
$q$ for these transitions is very large, $q \sim -10^5$ and
corresponding states are metastable. 

%########################################################################
\begin{acknowledgments}
This work was supported by the Australian Research Council
and the NCI National Facility.
\end{acknowledgments}

\end{document}